\documentclass[]{aastex631}
\usepackage{graphicx}
\usepackage{amsmath}
\usepackage{mathtools}
\usepackage{amssymb}
\DeclareMathOperator{\sech}{sech}
\shorttitle{Slow Shock Formation Upstream of Reconnecting Current Sheet}
\shortauthors{Arnold et al.}
\graphicspath{{./}{figures/}}

\begin{document}

\title{Slow Shock Formation Upstream of Reconnecting Current Sheets}

\author{H.~Arnold}
\affiliation{IREAP, University of Maryland, College Park MD 20742-3511, USA}
\author{J.~F.~Drake}
\affiliation{IREAP, University of Maryland, College Park MD 20742-3511, USA}
\author{M.~Swisdak}
\affiliation{IREAP, University of Maryland, College Park MD 20742-3511, USA}
\author{F.~Guo}
\affiliation{Los Alamos National Laboratory, Los Alamos NM 87545, USA}
\author{J.~T.~Dahlin}
\affiliation{NASA Goddard Space Flight Center, Greenbelt MD 20771, USA}
\author{Q. Zhang}
\affiliation{Los Alamos National Laboratory, Los Alamos NM 87545, USA}


\begin{abstract}
The formation, development and impact of slow shocks in the upstream regions of reconnecting current layers are explored. Slow shocks have been documented in the upstream regions of magnetohydrodynamic (MHD) simulations of magnetic reconnection as well as in similar simulations with the {\it kglobal} kinetic macroscale simulation model. They are therefore a candidate mechanism for preheating the plasma that is injected into the current layers that facilitate magnetic energy release in solar flares. Of particular interest is their potential role in producing the hot thermal component of electrons in flares. During multi-island reconnection, the formation and merging of flux ropes in the reconnecting current layer drives plasma flows and pressure disturbances in the upstream region. These pressure disturbances steepen into slow shocks that propagate along the reconnecting component of the magnetic field and satisfy the expected Rankine-Hugoniot jump conditions. Plasma heating arises from both compression across the shock and the parallel electric field that develops to maintain charge neutrality in a kinetic system. Shocks are weaker at lower plasma $\beta $, where shock steepening is slow. While these upstream slow shocks are intrinsic to the dynamics of multi-island reconnection, their contribution to electron heating remains relatively minor compared with that from Fermi reflection and the parallel electric fields that bound the reconnection outflow. 
\end{abstract}



\section{Introduction} \label{sec:intro}
Magnetic reconnection is the driver of explosive phenomena throughout the the universe. Within the heliosphere it is the mechanism by which magnetic energy is converted into plasma energy in solar flares and in the Earth's magnetosphere. Plasma energy takes the form of bulk flow energy, thermal energy and  nonthermal energy, which typically manifests as powerlaw energy distributions. While the basic physics of reconnection, including the factors that control the rate of reconnection, are mostly understood \citep{Shay1998,Hesse1999,Shay2007,Liu2017}, significant issues remain unresolved -- especially those related to particle energy gain and the partitioning of energy between thermal and non-thermal particles. X-ray observations of flares, for example, reveal photon spectra that take the form of thermal distributions at low energy that roll into powerlaw distributions at energies above 5-10 keV \citep{Lin2003,Emslie2012}. The powerlaw distributions can extend to 10's of MeV in energetic events \citep{Vilmer2012}. Broad surveys of electron heating in flares are consistent with this picture. A study of 380 M- and X-class flares revealed that the thermal electron temperature was around 8-10 MK, significantly higher than the ambient coronal temperature of $\sim 1-4$ MK \citep{Aschwanden2016}. Some of the increase in thermal energy can be attributed to the evaporation of chromospheric plasma due to energetic electron deposition. However, the observation of such hot thermal distributions high in the corona suggests that reconnection can drive the hot thermal electrons \citep{Krucker2010,Oka2013}. Observations of ``hot onset" in flares, in which the hot electrons are inferred before the onset of HXRs \citep{Hudson2021}, further suggest that reconnection-driven electron thermal heating is active.

In a study of 79 reconnection events in the Earth's magnetosphere, \cite{Phan2013} found that the electron heating is proportional to the available magnetic energy per particle, $m_iC_{Ar}^2$ where $m_i$ is the ion mass and $C_{Ar}$ is the Alfv\'en speed based on the reconnecting field. Observations of reconnection in the Earth's magnetotail, in which $m_iC_{Ar}^2$ is comparable to that in flares, reveal thermal electron distributions that roll into powerlaw tails \citep{Oieroset2002,Ergun2020} as in flare observations. PIC simulations reproduce the scaling of energy gain by electrons seen in the magnetosphere observations \citep{Haggerty2015}. While slow shocks have been widely invoked to explain flare observations \citep{Tsuneta1996,Longcope2010}, PIC simulations suggest that the slow shocks in flares remain laminar and do not produce significant electron energy gain, especially with a low upstream plasma $\beta=8\pi P_{plasma}/B^2$ \citep{Zhang2019a, Zhang2019b}, where $P_{plasma}$ is the plasma pressure and $B$ is the magnetic field. Instead, electron heating is dominated by the development of a weak potential that bounds the reconnection exhaust to maintain charge neutrality \citep{Zhang2019a, Zhang2019b}. The absence of powerlaw tails in these PIC simulations means the mechanisms that energize the hot thermal, in contrast with the non-thermal, component have not been identified. 

Simulations with the macro-scale simulation model {\it kglobal} revealed the formation of hot thermal as well as non-thermal electrons in a reconnecting current sheet \citep{Arnold2021}. Fermi reflection dominated the energy gain of the non-thermal electrons and was found to be sensitive to the ambient guide (out-of-plane) magnetic field. In contrast, the  thermal electron heating scaled as $\sim 0.04m_iC_A^2$ and was insensitive to the guide field.
These simulations also revealed that large numbers of slow shocks formed in the inflow region upstream of the reconnecting current sheet. We suggested that the potential drop across the slow shock associated with electron pressure gradients could heat the thermal electrons in the upstream region.  Heating by slow shocks has been documented in measurements in the Earth's magnetotail \citep{Schwartz1987}. Once heated in the upstream region,  electrons would then advect into the reconnecting current sheet and then into the plasmoids that populate the current sheet, where the thermal heating was measured \citep{Arnold2021}. Underlying this suggestion was the idea that such a mechanism would be insensitive to the ambient guide field, as was suggested by the simulations \citep{Arnold2021}. While the formation of slow shocks upstream of reconnecting current sheets has not been widely recognized, previous MHD simulations revealed the formation of these structures \citep{Karpen2012, Zenitani2015} and led to the idea that the plasma compressive motion in the upstream region could increase the local plasma pressure \citep{Zenitani2020}. 

Due to the potential influence that upstream slow shocks might have on electron heating, it is important to investigate both how they form and their consequences. The earlier MHD simulations suggested that the slow shocks form due to the motion of plasmoids within the reconnecting current sheets although the details of slow shock formation under these conditions were not entirely clear. In particular, if the shocks form due to plasmoids exceeding the slow mode speed, then they will form on the front of the plasmoid and move at the same speed as the plasmoid (e.g., a bow shock). However, as we show below, the slow shocks observed in large scale simulations form due to wave steepening instead, and therefore move independent of the plasmoid.

In this paper we explore the mechanism for slow shock formation in the upstream region and explore the consequences of their formation for electron heating. In Section \ref{numerical setup} we describe the simulations used for this study. In Section \ref{slow shock formation} we describe how the shocks are formed. We discuss the electron heating due to the slow shocks in Section \ref{ssheating}. Finally, in Section \ref{conclusion} we discuss how the shock formation changes based on the strength of the guide field and the upstream value of $\beta$ based on the reconnecting magnetic field, and implications for electron heating in the early stages of solar flares.

\section{Numerical Setup \label{numerical setup} }
Using the {\it kglobal} model, we carried out 2D simulations. {\it kglobal} uses a magnetohydrodynamic
(MHD) backbone that contains fluid ions but also has macroparticle electrons that feed back on the MHD fluid. In order to preserve charge neutrality a fluid electron species is also included.
\citep{Drake2019,Arnold2019}.
The upstream reconnection magnetic field, $B_0$, and the ion density,
$n_0$, define the Alfv\'en speed, $C_{A0}=B_0/\sqrt{4\pi
  m_in_0}$. Lengths are normalized to an arbitrary length $L_0$, since there are no inherent length scales. Times are normalized to an Alfve\'nic time scale,
$\tau_A=L_0/C_{A0}$. Temperatures and particle energies are normalized to
$m_iC_{A0}^2$. Following an MHD scaling, the perpendicular electric field scales like $C_{A0}B_0/c$,where $c$ is the speed of light. From \cite{Arnold2019} the parallel electric field is:
\begin{equation}
    E_{\parallel}=\frac{-1}{n_ie}\left( \boldsymbol{B} \cdot \boldsymbol{\nabla} \left( \frac{m_en_f v_{\parallel f}^2}{B} \right) + \boldsymbol{b}\cdot \boldsymbol{\nabla} P_{f} + \boldsymbol{b} \cdot \boldsymbol{\nabla} \cdot \boldsymbol{T}_h\right)
    \label{Epar}
\end{equation}
where $m_e$ is the electron mass, $n_i$ and $n_f$ are the ion and fluid electron densities respectively, $P_f$ is the fluid electron pressure, $\boldsymbol{T_h}$ is the gyrotropic particle electron stress tensor, $v_{||f}$ is the component of the fluid electron velocity parallel to the magnetic field, $B$ is the magnetic field, and $e$ is the elementary charge. From Eq. \ref{Epar} and by noting that the pressure terms scale like $m_i n_i C_{A0}^2$, it is clear that $E_{||}$ scales like $m_i
C_{A0}^2/eL_0$, and is therefore smaller than the perpendicular
component by the ion inertial length divided by $L_0$. Importantly, a parallel potential drop over $L_0$ can have an energy that is of the order $m_iC_{A0}^2$, which, during magnetic reconnection, is
comparable to the available magnetic energy per particle.

The simulations are initialized with constant densities and pressures
in two force-free current sheets with periodic boundary conditions. For simplicity we will usually just focus on one current sheet. Thus,
$\boldsymbol{B}=B_0 \tanh{(y/w)}\boldsymbol{\hat{x}}+ \sqrt{B_0^2
  \sech^2{(y/w)}+B_g^2} \boldsymbol{\hat{z}}$, where $w$ is a constant controlling the initial width of the current sheet. The temperatures of all
three species are uniform and isotropic with
$T_i=T_{e,part}=T_{e,fluid}$. The chosen
fraction of particle electrons is $25\%$ of the total electrons with the remaining $75\%$ represented as a fluid. The
domain size for all simulations is $L_x \times L_y = 2\pi L_0 \times
\pi L_0$ including both current sheets. The magnetic field evolution equation includes a
hyper-resistivity $\nu$ to facilitate reconnection, while minimizing
dissipation at large scales \citep{Kaw1979}. This defines the effective Lundquist
number $S_\nu=C_AL_0^3/\nu$ associated with the hyper-resistivity and at the same time defines the effective system size -- a larger value of $S_\nu$ produces more magnetic flux ropes as reconnection proceeds. 
We include diffusion on the particle electrons in order to prevent a numerical instability where fluctuations in the perpendicular electric field can trap particles. Additionally fourth- and second-order viscosity terms are included to kill instabilities that grow at the grid scale. The particle noise causes reconnection to begin and produce multiple flux ropes whose number increases
with $S_\nu$. Unless
otherwise stated we focus on simulations with 100 particles per cell,
time step $dt=0.0005 \tau_0$, $S_\nu=9.5\times 10^7$, and a uniform grid of $N_x\times
N_y=2048 \times 1024$ cells. If $L_0=10^4km$, a reasonable macro-scale in the corona, our grid cell
is 30 km across, much larger than any kinetic scale or the entire domain in computationally feasible particle-in-cell (PIC)
simulations given that the Debye length is $\sim 1cm$. The mass ratio is $m_i/m_e=100$. The speed of light is $c/C_{A0} \approx
60$. We primarily consider simulations with guide fields of $B_g/B_0=0.25$ and $0.6$ and $\beta_r=0.5$ and $0.0625$.

\section{Slow Shock Formation \label{slow shock formation}}
\begin{figure}
\centering
\includegraphics[width=32pc,height=22pc,keepaspectratio]{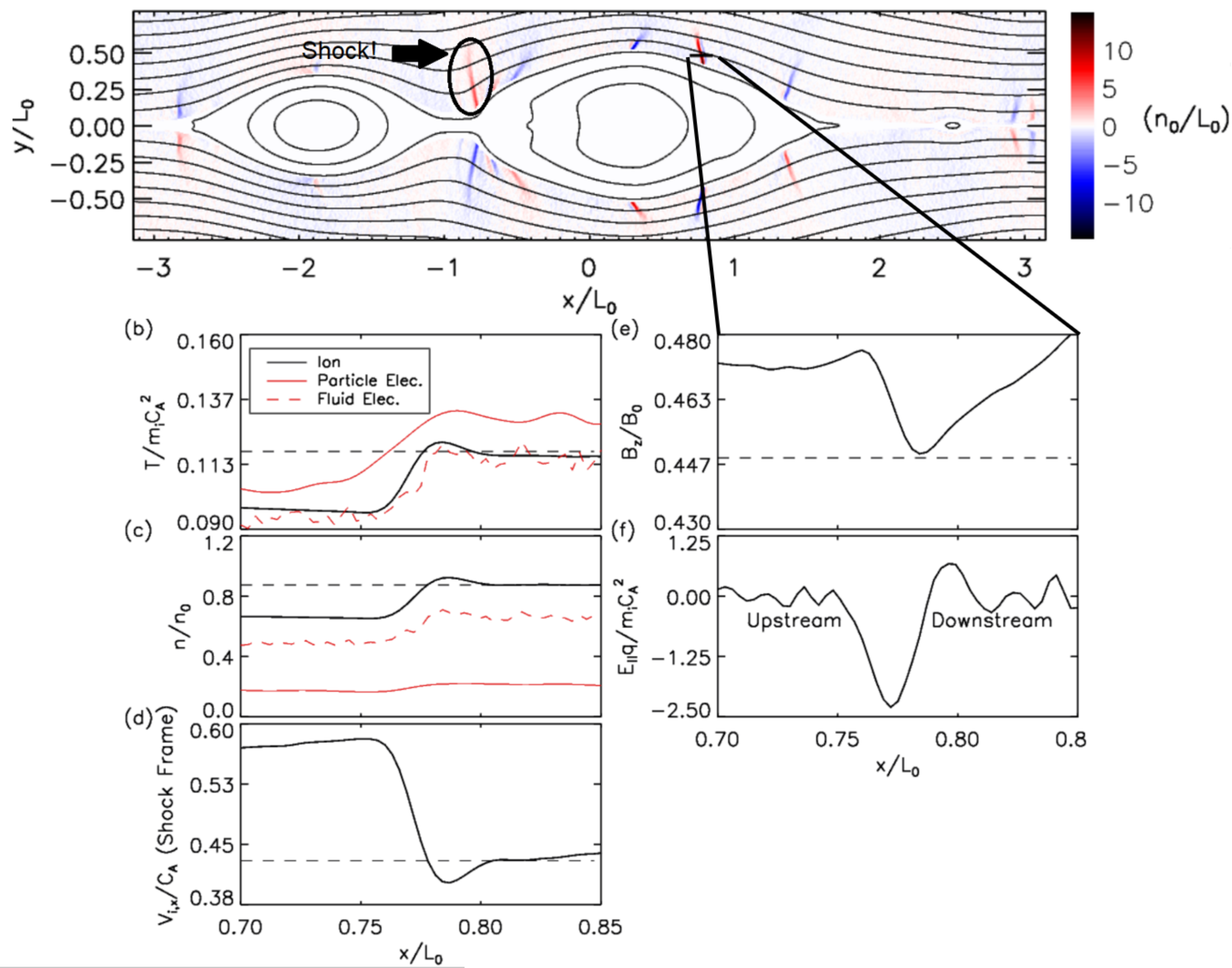}
\caption{Overview of the formation of slow shocks upstream of a reconnecting current sheet from a {\it kglobal} simulation with $B_g/B_0=0.6$ and $\beta_r = 0.5$. In panel (a): $\boldsymbol{\hat{b}} \cdot \boldsymbol{\nabla} n_i$ with overlaid magnetic field lines and the regions inside of the outermost separatrix zeroed out to emphasize the upstream region at $t/\tau=14$. The magnetic field points to the right above the current sheet. One shock is circled for clarity. Another shock around $x=0.7L_0$, $y=0.5L_0$ is traveling at a speed of $\sim -0.47 C_A0$ in the x-direction and is cut by a black line along which various quantities are plotted in panels (b)-(f). These cuts are respectively the temperature of each species, the density of each species, the ion flow speed in the x-direction in the shock frame, the guide field, and the parallel electric field. The black dashed horizontal lines correspond to the respective ion quantity in the downstream region according to the Rankine-Hugoniot jump conditions.}
\label{shock}
\end{figure}

\begin{figure}
\centering
\includegraphics[width=32pc,height=22pc,keepaspectratio]{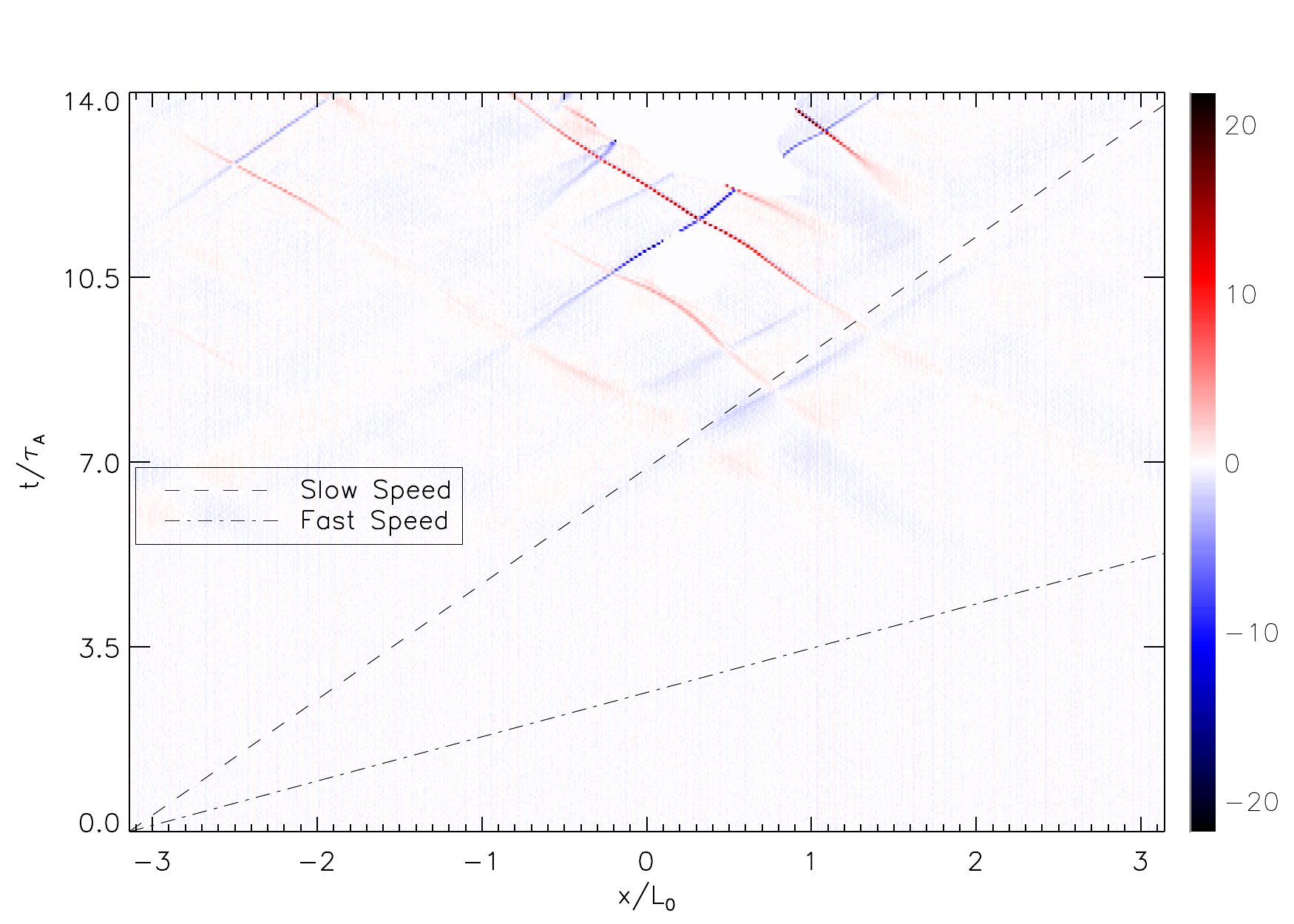}
\caption{A time space plot of $\boldsymbol{\hat{b}} \cdot \boldsymbol{\nabla} n_i$ taken along a horizontal cut at $y/L_0=0.38$ in Fig.~\ref{shock}. Lines corresponding to the slow (dashed) and fast (dot-dashed) speeds are overplotted for reference. Note that since the shocks travel along the reconnecting field lines the fast speed in the presence of a guide field is $v_f=\sqrt{\frac{1}{2}(c_A^2+c_s^2+\sqrt{(c_A^2+c_s^2)^2-4c_{A,r}^2c_s^2})} \approx 1.19 c_{A0}$ where $C_{A,r}$ is the Alfv\'en speed based solely on the reconnecting field.}
\label{timespace}
\end{figure}

 Fig.~\ref{shock} shows key properties of our simulation with $B_g/B_0=0.6$ and $\beta=0.5$. Shown in Fig. \ref{shock}(a) is the parallel gradient of the ion density overlaid with magnetic field lines. The shocks can be identified as the red and blue structures in the upstream, inflow, region that are primarily vertical. One is circled for clarity. At any given time, numerous shocks travel to the right and left. The speed of these shocks measured in the lab frame is approximately the slow-mode speed as can be seen in Fig.~\ref{timespace} which depicts a time space plot of Fig.~\ref{shock}(a)  where the slow-mode speed is overplotted as a dashed line for reference. In Figs.~\ref{shock}(b)-(c) the temperature, $T$, and density, $n$, of the three species are shown in a horizontal cut across the shock centered around $[x, y]=[0.7L_0, 0.5L_0]$. The ion velocity in the shock frame $V_{ix}$, the parallel electric field $E_\parallel$, and $B_z$ are shown in (d)-(f). Unlike in fast mode shocks, the magnetic field downstream of slow shocks decreases, as is evident in Fig.~\ref{shock}(e). The upstream Mach number $M$ of the shock corresponding to the 
 cuts is $1.23$, where the Mach number is defined as 
 \begin{equation}
     M^2=\frac{v_{xu}^2(1+B_{zu}^2/B_x^2)}{C_{su}^2}
     \label{machnumber}
 \end{equation}
 where the $u$ ($d$) subscript corresponds to upstream (downstream) quantities, $v_{xu}$ is the value of $V_{ix}$ in the upstream region, and $C_{su}=\sqrt{5P_u/3\rho_u}$ is the sound speed. Note that in the de Hoffman-Teller frame $v_{zu}=v_{xu}B_{zu}/B_x$ so the $M$ in Eq.~(\ref{machnumber}) is defined with the total parallel velocity where we have neglected any component in the $y$ direction since $B_y \approx 0$. In terms of $M$, the compression ratio $R=n_d/n_u$ across the shock is given by 
 \begin{equation}
     R=\frac{4M^2}{M^2+3}.
     \label{R}
 \end{equation}
 with the downstream velocity $v_{xd}=v_{xu}/R$, the downstream temperature
 \begin{equation}
     \frac{T_d}{T_u}=1+\frac{M^2}{3}\left( 1-\frac{1}{R^2}\right)
     \label{Rtemp}
 \end{equation}
 and the downstream magnetic field
 \begin{equation}
     \frac{B_{zd}}{B_{zu}}=1+M^2\frac{C_{su}^2}{C_{Au}^2}\left( \frac{1}{R}-1\right)
     \label{RBz}
 \end{equation}
 where $C_{Au}=(C_{Axu}^2+C_{Azu}^2)^{1/2}$ is the Alfv\'en speed based on the total magnetic field. Since $C_{su}^2/C_{Au}^2\ll 1$ the decrease in $B_z$ across the shock is small. The expected downstream values of the various parameters across the shock in Fig.~\ref{shock}(b)-(e) are shown by the dashed horizontal lines. Deviations from  Rankine–Hugoniot jump conditions \citep{Lin1993b} are surprisingly small which suggests that the particle electrons are not playing a major role in controlling the shock structure. 
 
\begin{figure}
\centering
\includegraphics[width=32pc,height=22pc,keepaspectratio]{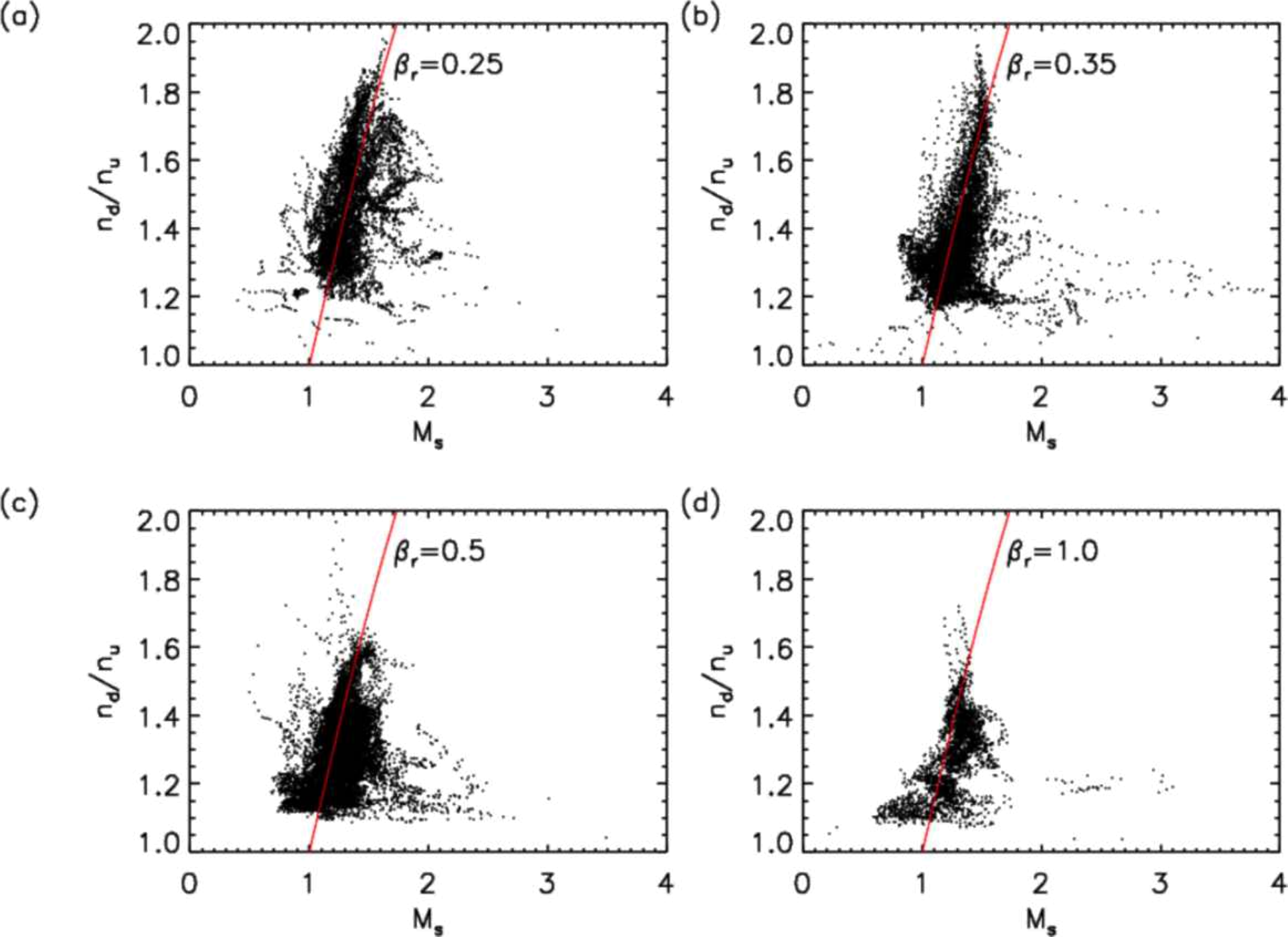}
\caption{Slow shock compression versus Mach number for a variety of values of $\beta_r$. The red lines are the compression ratio expected from the MHD jump conditions. Note that these simulations have $B_g/B_0=0.4$, $m_i/m_e=25$, and a domain half the size of the other simulations presented in this paper.}
\label{shockjumps}
\end{figure}

 Shocks also form with a variety of initial guide fields and $\beta$. Shown in Fig.~\ref{shockjumps} are the measured compression ratios versus the Mach number for shocks from simulations with four values of $\beta_r$. The measured Mach numbers cluster between one and two. The red lines are the predictions from the MHD jump condition given in Eq.~(\ref{R}). The measured compression ratios generally follow the trend given by the MHD jump conditions, confirming that these are indeed slow shocks. The jump conditions have been evaluated at many values of $y$ along the extension of the shock in this direction, causing each shock to be represented as multiple data points. The substantial scatter of the data in the figure is a consequence of noise in the simulation data and the challenge of identifying upstream and downstream parameters for discontinuities that are at or marginally above the $M_s>1$ shock threshold. 

The upstream slow shocks that developed in earlier MHD simulations were linked to the motion of plasmoids in reconnecting current layers \citep{Zenitani2015,Zenitani2020}. However, the specific mechanism driving these shocks was not discussed in detail.
Since slow shocks are associated with the compressional motion of plasma parallel to the ambient magnetic field, it is the parallel forces acting on ions that triggers shock formation. By dotting Equation (5) in \cite{Arnold2019} with $\boldsymbol{b}$ and solving for the time derivative of the parallel ion flux, $n_iv_{i\parallel}$, we arrive at equation \ref{parflux} which shows the four drivers of said flux in the {\it kglobal} model,
\begin{equation}
    \frac{\partial m_in_iv_{i,||}}{\partial t} =
    \frac{n_im_i}{2} \boldsymbol{\hat{b}} \cdot \boldsymbol{\nabla} v_{\perp}^2+
    n_im_iv_{i,||} \boldsymbol{v_{\perp}} \cdot \boldsymbol{\kappa}-
    \boldsymbol{\nabla} \cdot \boldsymbol{v_i} m_in_iv_{i,||}-
    \boldsymbol{\hat{b}} \cdot \boldsymbol{\nabla} (P_i+P_{e,f} +P_{e,p})
    \label{parflux}
\end{equation}
where $m_i$ is the ion mass, $n_i$ is the ion density, $v_{i,||}$ is the parallel ion flow speed, $\boldsymbol{\hat{b}}$ is the magnetic unit vector, $v_\perp$ is the perpendicular ion flow speed, $\boldsymbol{\kappa}$ is the magnetic curvature, $P_i$, $P_{e,f}$ and $P_{e,p}$ are the ion, electron fluid and electron particle pressures, respectively. For simplicity, we have ignored the anisotropy in the electron pressure in the upstream region. Thus, Eq.~(\ref{parflux}) in the {\it kglobal} model is unchanged from that in the conventional MHD treatment. 

\begin{figure}
\centering
\includegraphics[width=32pc,height=22pc]{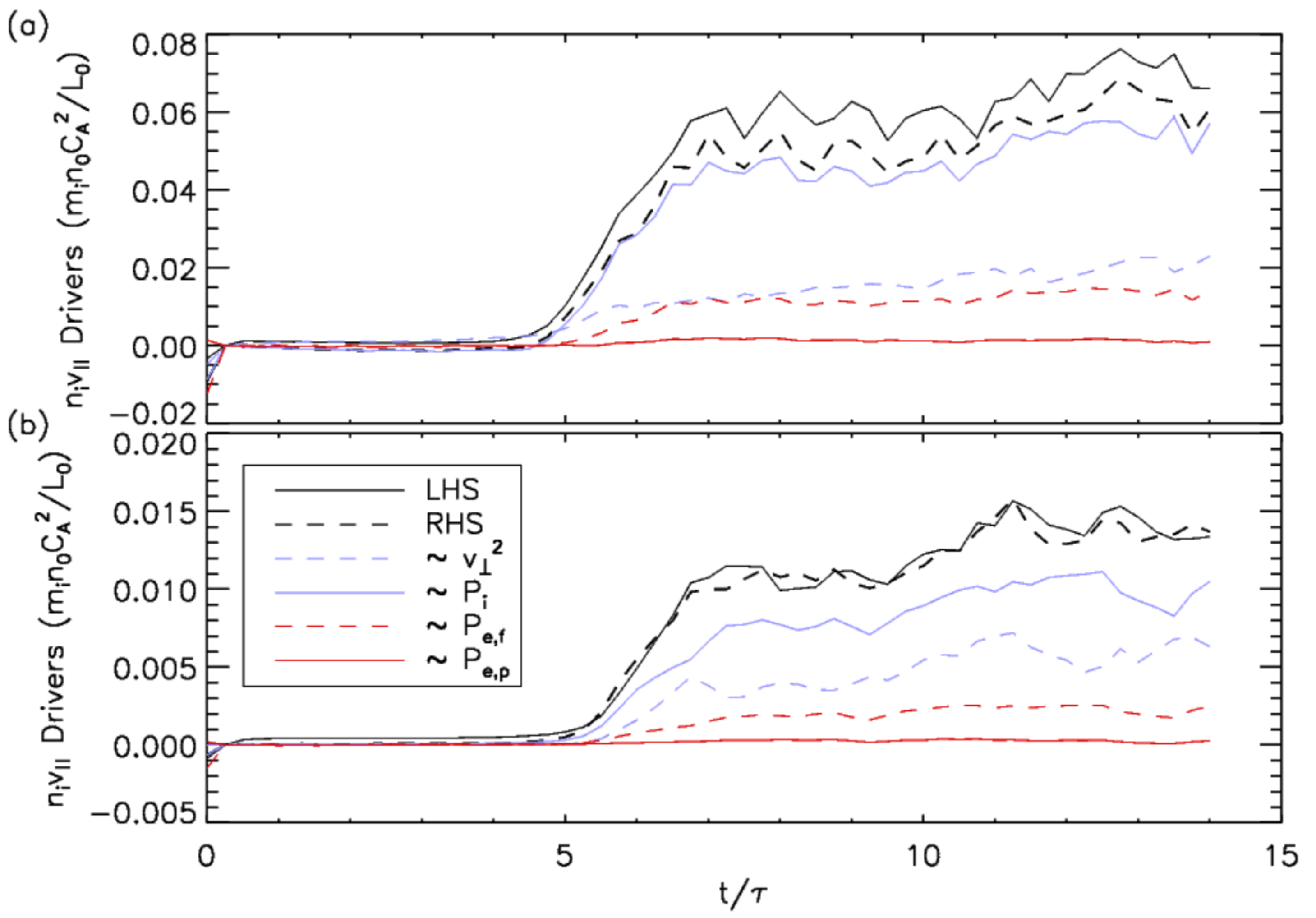}
\caption{The drive mechanism of plasma flows leading to slow shock formation upstream of the reconnecting current layer. Shown are the integrated values of the smoothed drive terms for the parallel ion flow in  Eq.~\ref{parflux} as a function of time for $\beta_r=0.5$ in panel (a) and $\beta_r=0.0625 $ in panel (b). Both simulations have $B_g/B_0=0.6$.}
\label{jpardrive}
\end{figure}

The middle two terms on the right-hand side of Eq.~(\ref{parflux}) are proportional to $v_{i\parallel}$ and therefore cannot be the initial drivers of parallel flow. Thus, to drive parallel flow in the upstream region, either the first term or the three pressure terms must be responsible. To isolate which terms drive the initial flows, and therefore lead to the slow shocks, we first smooth each term to reduce noise, then integrate the absolute value of each term over the region upstream of the outermost separatrix. The result is plotted in Fig.~\ref{jpardrive} as a function of time for two different values of $\beta _r$. In the case of $\beta _r=0.5$ the ion pressure is the dominant driver of the parallel ion flow. Even in the run with $\beta_r=0.0625$ where the $V_\perp^2$ term becomes comparable to the pressure, it is still consistently smaller and increases later in time than the pressure. Thus, the overall dominance of the ion pressure term in the parallel flow driver when the upstream flow is forming suggests that the slow shocks develop from parallel pressure disturbances that then steepen into shocks. 

Because all of the four driving terms for $V_{i\parallel}$ in Eq.~(\ref{parflux}) are proportional to the parallel gradient of either the pressure or the perpendicular flow energy (excluding those proportional to $V_{i,\parallel}$), the guide field will act to reduce these terms since there is no variation in the out-of-plane direction. This in turn will cause the nonlinear terms to be smaller and suppress the steepening into shocks. Notably, the observed flows in the upstream region are not superslow, i.e. faster than the slow speed, in the ``lab frame", but result in typical Mach numbers of $\sim 1.2-1.6$ in the frame of the shock. This is consistent with simulations of different guide fields and plasma $\beta$ as can be seen in Fig. \ref{shockjumps}.

In current layers undergoing multi x-line reconnection, plasmoids grow and merge on Alfv\'enic time scales. Their motion along the current sheet, driven by outflows from x-lines and merging, creates pressure changes in front and behind them in the upstream region. These pressure gradients then drive the parallel flows that steepen into slow shocks. Importantly, the plasmoids rarely reach Alfv\'enic speeds, despite Alfv\'enic flows, since the merging process effectively decelerates them. By considering a plasmoid merger as an elastic collision it is clear that the center-of-mass momentum is conserved, whereas the individual plasmoid momentum is not. Thus the merging process inhibits plasmoids from reaching superslow speeds. This is why we do not observe shocks created by superslow motion in a current layer undergoing multi x-line reconnection.

\begin{figure}
\centering
\includegraphics[width=32pc,height=22pc]{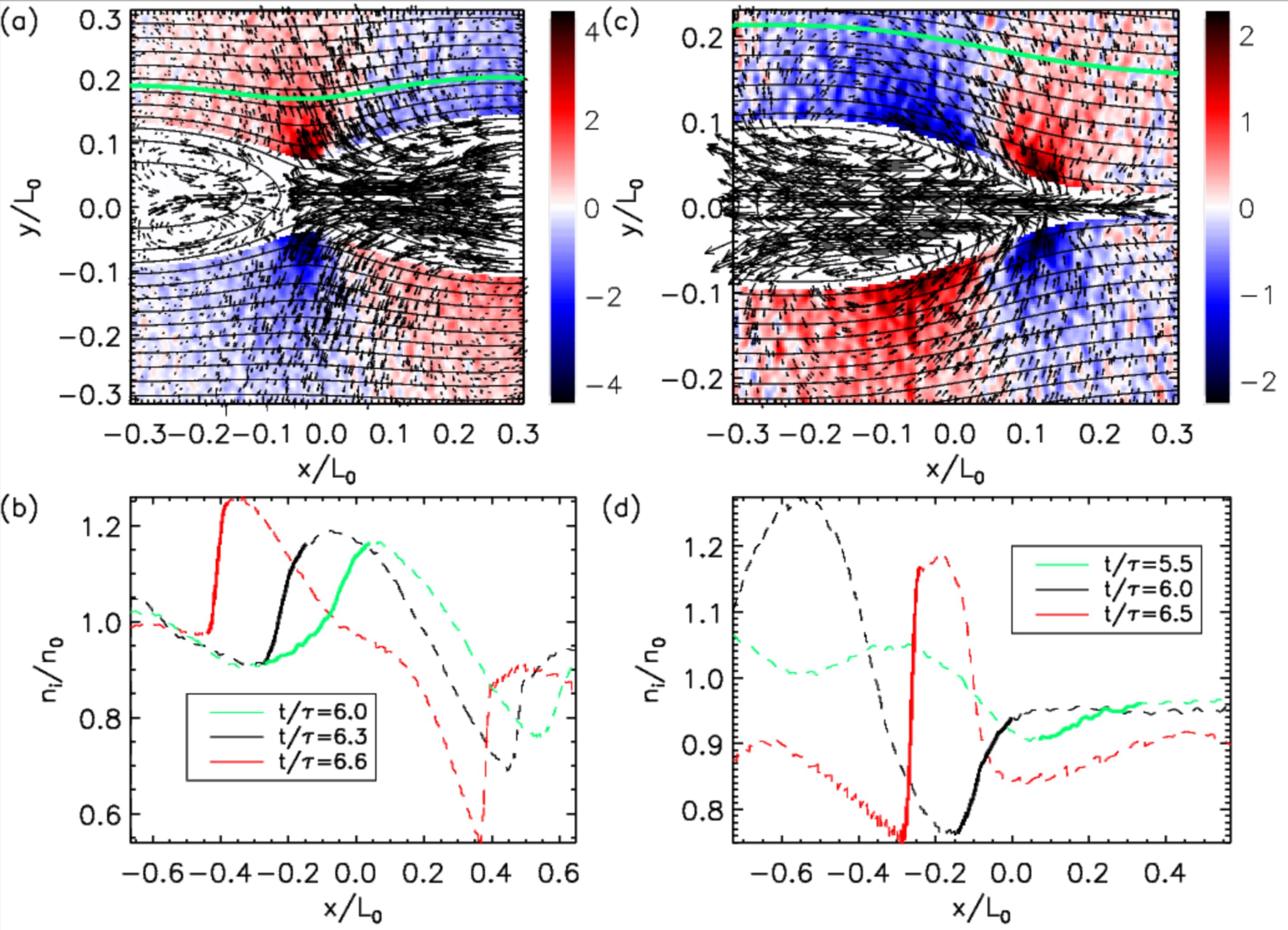}
\caption{Two mechanisms for the generation of pressure disturbances upstream of the reconnecting current sheet resulting from flux rope merging in (a)  and convection in (c). In panels (a) and (c): $\boldsymbol{\hat{b}} \cdot \boldsymbol{\nabla} n_i$ with overlaid magnetic field lines and arrows representing the ion flows with the regions inside of the outermost separatrix zeroed out to emphasize the upstream region. Panel (a) is at $t/\tau=6.0$ and panel (b) is at $t/\tau=5.5$ which matches the green lines in panels (b) and (d) respectively. The magnetic field points to the right above the current sheet in both images. Note that the axes are shifted such that (0,0) is the center of the image for both panels. In Panels (b) and (d): the ion density along the green field line in panels (a) and (c) for three different times. The thick and solid sections of each line show the approximate pressure gradient region. The initial parameters are $B_g/B_0=0.6$ and $\beta_r=0.5$. Note the existence of a second shock at the final time in panel (b) around $x/L_0=0.35 - 0.4$. The axes in panels (b) and (d) are longer along the x-direction to add context, but are centered to match the axes in (a) and (c) respectively.}
\label{shockflow}
\end{figure}

In Fig. \ref{shockflow} we show two generic shock formation mechanisms associated with the motion of flux ropes. In Panel (a) two islands collide with each other. This forces the plasma in the upstream region to flow away from the magnetic separatrix and creates a pressure pileup around $x/L_0=0$. The pressure gradient along a particular field line then steepens to form a shock if the field line remains in the upstream region long enough before flowing into the current layer and reconnecting. The density profile along the green magnetic field line in Panel (a) is shown at three times in Fig.~\ref{shockflow}(b).  At $t/\tau_A=6.0$ there is a clear gradient, which steepens into a shock around $x/L_0=-0.5$ by $t/\tau_A=6.6$. 

A similar but distinct mechanism for shock formation is shown in Fig.~\ref{shockflow}(c). In this case, the island is far from other islands but is moving with high velocity to the left. Its motion leads to a pressure drop in its wake. As before (see Fig.~\ref{shockflow}(d)), the pressure gradient steepens into a shock. However, in this case the upstream flows are directed towards the current sheet. Because the upstream flow is towards the current sheet the time available for the pressure disturbance to steepen into a shock before being injected into the current sheet is reduced. The shock formation time $\tau_s$ should scale like the parallel scale length of the pressure disturbance, which is of the order of the size of the flux rope $L_{fr}$ divided by the sound speed so that $\tau_s\sim L_{fr}/C_s$. Thus in systems with low upstream $\beta$ the steepening process is slow compared with the inflow speed which scales with the Alfv\'en speed. Thus, there is insufficient time for pressure disturbances upstream to steepen into shocks before flowing into the reconnecting current layer. An important consequence of these shock formation mechanisms is that the upstream flows, and thus the resulting shocks, will have vertical spatial extents that are comparable to the width of the magnetic islands in the region upstream of the current sheet.

\section{Electron heating from upstream slow shocks}\label{ssheating}
\begin{figure}
\centering
\includegraphics[width=32pc,height=22pc]{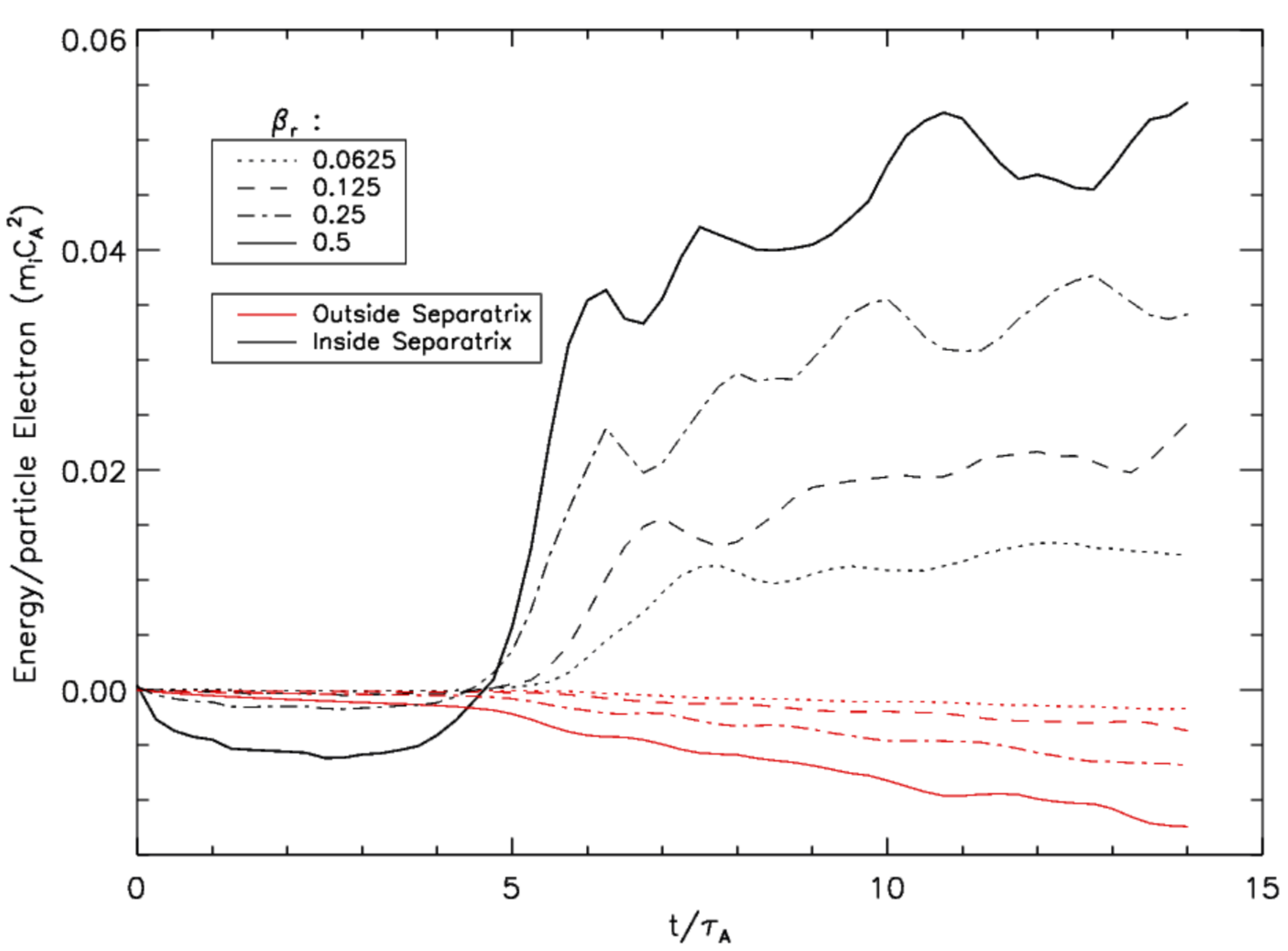}
\caption{The energy gain per particle electron is shown as a function of time for four different values of $\beta_r$ both inside (black) and outside (red) the outermost separatrix.}
\label{enden}
\end{figure}

An important question is whether the compressive motion of the plasma upstream of the reconnecting current sheet and the associated slow shocks can play a significant role in heating the plasma before it directly interacts with plasmoids in the current sheet. In the earlier {\it kglobal} reconnection simulations the temperature increment of the hot thermal electrons was found to be insensitive to the ambient guide field, suggesting a heating mechanism other than Fermi reflection was the driver for the heating seen \citep{Arnold2021}. Thus, slow shocks in the upstream region were proposed as a possible heating mechanism of the hot thermal electrons \citep{Zenitani2020,Arnold2021}. Shown in Fig.~\ref{enden} is the average energy change per particle of the particle electrons in the upstream (red) and downstream (black) regions as a function of time for several values of $\beta_r$ for $B_g/B_0=0.6$. While significant energy gain is observed inside the separatrix as expected, the upstream region actually experiences net cooling. The net cooling is likely due to a combination of upstream expansion and a reverse version of Fermi reflection since $\boldsymbol{v}_i \cdot \boldsymbol{\kappa} < 0$. If the slow shocks were responsible for the heating of the thermal electrons, then we would expect to see some energy gain in the upstream region. Additionally, while Fig. \ref{shock} (b) shows that the slow shocks do heat the electrons, the downstream temperature is roughly equal to the initial upstream temperature $T_0=0.125$. Therefore, the slow shocks in the present simulations can not be the primary mechanism by which the thermal electrons gain energy. The mechanism driving thermal energy gain therefore remains an open question.

\section{Conclusion \label{conclusion}}
An important observation in flares and during magnetic reconnection events in the solar wind and the Earth's magnetosphere is the formation of a hot thermal component of the electron energy distribution. Slow shocks develop in the upstream region of both MHD \citep{Zenitani2015,Zenitani2020} and {\it kglobal} simulations of reconnecting current layers. The mechanism for the formation of these slow shocks has been explored here with {\it kglobal} simulations. The motion of plasmoids sets up pressure perturbations well upstream of the reconnecting current layer. These disturbances steepen into shocks. In low-$\beta_r$ systems we expect there to be relatively fewer shocks since the shock formation time scales inversely with the sound speed while the time scale for convection into the reconnecting current sheet is linked to the Alfv\'en time. The consequence is that slow shock formation is limited to cases when the upstream flow generated by plasmoid motion is directed away from the current sheet. This takes place when plasmoids converge. We can estimate the $\beta_r$ below which slow shock formation is reduced by setting the sound speed equal to the reconnection inflow speed. This yields $\beta_r\sim 0.01$ for a rate of reconnection that is of the order of $0.1C_{Ar}$. However, this is likely a low estimate since as shown in Fig.~\ref{shockflow}(c) the inflow speed along the edge of a plasmoid can be greater than the typical reconnection inflow speed. 

Although the conclusion based on simple slab geometry simulations of a reconnecting current sheet is that the upstream slow shocks are too weak to drive significant electron heating upstream of the current layer, this may not be the case for a realistic solar flare configuration that results in the launch of a large-scale CME. Since the size of the slow shocks scales with the plasmoid, it is possible that the high-velocity launch of a CME could cause similarly sized downflows containing a large amount of free energy. Simulations with more realistic solar flare geometries, and upstream resolution sufficient to resolve shocks, are needed to be able to determine the contribution to thermal heating due to the slow shocks described in this paper.

While slow shocks are apparently regularly produced in macro-scale simulation models based on the MHD equations and those of {\it kglobal}, they have not yet been identified in PIC models of reconnection. One possibility is that sound waves are typically damped in plasma in which $T_e\sim T_i$ because the sound speed is comparable to the ion thermal speed. On the other hand, PIC simulations do reveal that large-amplitude sound waves do steepen into shocks in spite of their resonant interaction with ions \citep{Shay2007a}. 

\begin{acknowledgments}
The collaboration leading to these results was facilitated by the NASA Drive Science Center on Solar Flare Energy Release (SolFER), Grant No. 80NSSC20K0627. We would like to thank Dr. J. T. Karpen, Dr. C. R. DeVore, and participants in the NASA Drive Center SolFER for invaluable discussions that contributed to this work. This work has been supported by NSF Grants No. PHY1805829 and No. PHY1500460 and the FIELDS team of the Parker Solar Probe (NASA Contract No. NNN06AA01C) and the FINESST Grant No. 80NSSC19K1435. F. G. and Q. Z. acknowledges support in part from NASA Grant No. 80HQTR21T0104, 80HQTR21T0087, 80HQTR20T0073, 80HQTR20T0040, and DOE support through the LDRD program at LANL. J. T. Dahlin was supported by an appointment to the NASA Postdoctoral Program at the NASA Goddard Space Flight Center, administered by Universities Space Research Association under contract with NASA. The simulations were carried out at the National Energy Research Scientific Computing Center (NERSC). The data used to perform the analysis and construct the figures for this Letter are preserved at the NERSC High Performance Storage System and are
available upon request.  M.S. acknowledges support in part from NASA Grant No. 80NSSC20K1277. The authors appreciate the reviewer for their useful comments.
\end{acknowledgments}

\bibliography{lib}{}
\bibliographystyle{aasjournal}



\end{document}